**Sergey Yekimov**

Department of Trade and Finance, Faculty of Economics and Management, Czech University of Life Sciences Prague, Kamycka 129, 16500, Praha - Suchdol, Czech Republic

# SMOOTHING OF NUMERICAL SERIES BY THE TRIANGLE METHOD ON THE EXAMPLE OF HUNGARIAN GDP DATA 1992-2022 BASED ON APPROXIMATION BY SERIES OF EXPONENTS

## Abstract

In practice , quite often there is a need to describe the values set by means of a table in the form of some functional dependence . The observed values , due to certain circumstances , have an error . For approximation, it is advisable to use a functional dependence that would allow smoothing out the errors of the observation results.

Approximation allows you to determine intermediate values of functions that are not listed among the data in the observation table.

The use of exponential series for data approximation allows you to get a result no worse than from approximation by polynomials

In the economic scientific literature, approximation in the form of power functions, for example, the Cobb-Douglas function, has become widespread. The advantage of this type of approximation can be called a simple type of approximating function , and the disadvantage is that in nature not all processes can be described by power functions with a given accuracy.

An example is the GDP indicator for several decades . For this case , it is difficult to find a power function approximating a numerical series . But in this case, as shown in this article, you can use exponential series to approximate the data. In this paper, the time series of Hungary's GDP in the period from 1992 to 2022 was approximated by a series of thirty exponents of a complex variable.

The use of data smoothing by the method of triangles allows you to average the data and increase the accuracy of approximation . This is of practical importance if the observed random variable contains outliers that need to be smoothed out.

## Introduction

Data approximation is used to replace the observed numerical series with an analytical function that is as close as possible to the nodal points that correspond to the observed data. Currently, two main types of numerical series approximation have been put into practical use. By using an interpolation polynomial in the form of Newton or Lagrange, or by the least squares method.

The purpose of the approximating function is to smooth out random errors and distortions of data that occur during observations .

Measurement errors , as a rule , obey random laws , but observation errors do not always obey the normal distribution law . This can happen when the observed value depends on a large number of different factors.

In practice , many types of approximation have become widespread , among which one can distinguish:

1. Chebyshev approximation [1]

$$y^* = arg \min_{y} \max_{i=1,..,N} |f(t_i, y) - x_i|$$

2. Robust approximation [2]

$$y^* = arg \min_{y} \sum_{i=1}^{N} |f(t_1, y) - x_i|_0$$

where $|z|_0 = 1$, if $z \neq 0$, and $|z|_0 = 0$ if $z = 0$

3. Lp $(p \in [0, \infty))$ approximation [3]

$$y^* = arg \min_{y} \sqrt[p]{\sum_{i=1}^{N} |f(t_1, y) - x_i|^p}$$

4. L1 approximation [4]

$$y^* = arg \min_{y} \sum_{i=1}^{N} |f(t_1, y) - x_i|$$

5. Least squares approximation (Least Squares)[5]:

$$y^* = \arg\min_y \sum_{i=1}^{N}(f(t_1,y) - x_i)^2$$

5. Weighted least squares approximation (Weighted Least Squares)[6]:

$$y^* = \arg\min_y \sum_{i=1}^{N} \omega_i^2 (f(t_1,y) - x_i)^2$$

In the framework of this study, the author uses an approximation in the form of series of exponents. The possibility of approximation of numerical series by series of exponents is shown in the work [7]

## Methods

In carrying out this research, the author applied an analytical method by which the problems under study were considered in their unity and development. Taking into account the goals and objectives of the study, the structural and functional method of scientific research was used. This allowed us to study a number of problems related to the approximation of numerical series in the form of exponential series on the example of Hungary's GDP for 1992 – 2022.

## Results

One of the problems that researchers face when approximating numerical series is the presence of so-called outliers. Namely, if a hypothesis is put forward that a numerical series the value of a numerical series obeys a normal distribution, then the members of the numerical series whose magnitude differs significantly from other members may distort the final result. This may manifest itself in the fact that the arithmetic mean and the variance of the numerical series will differ from the result expected by the researcher. The solution to this problem may consist in the form of exclusion from the sequence of a numerical series of terms that are abnormally different from other terms.

The author proposes a procedure for smoothing numerical series by the triangle method. This procedure, according to the author, will allow to smooth out the outliers among the members of the numerical series.

Let's say there is tabular data
$(y_1, x_1), (y_2, x_2), (y_3, x_3), \dots, (y_{n-2}, x_{n-2}), (y_{n-1}, x_{n-1}), (y_n, x_n)$ they need to be approximated by some function $y = f(x)$ and let's say n we have equal 14 (Fig.1) , then

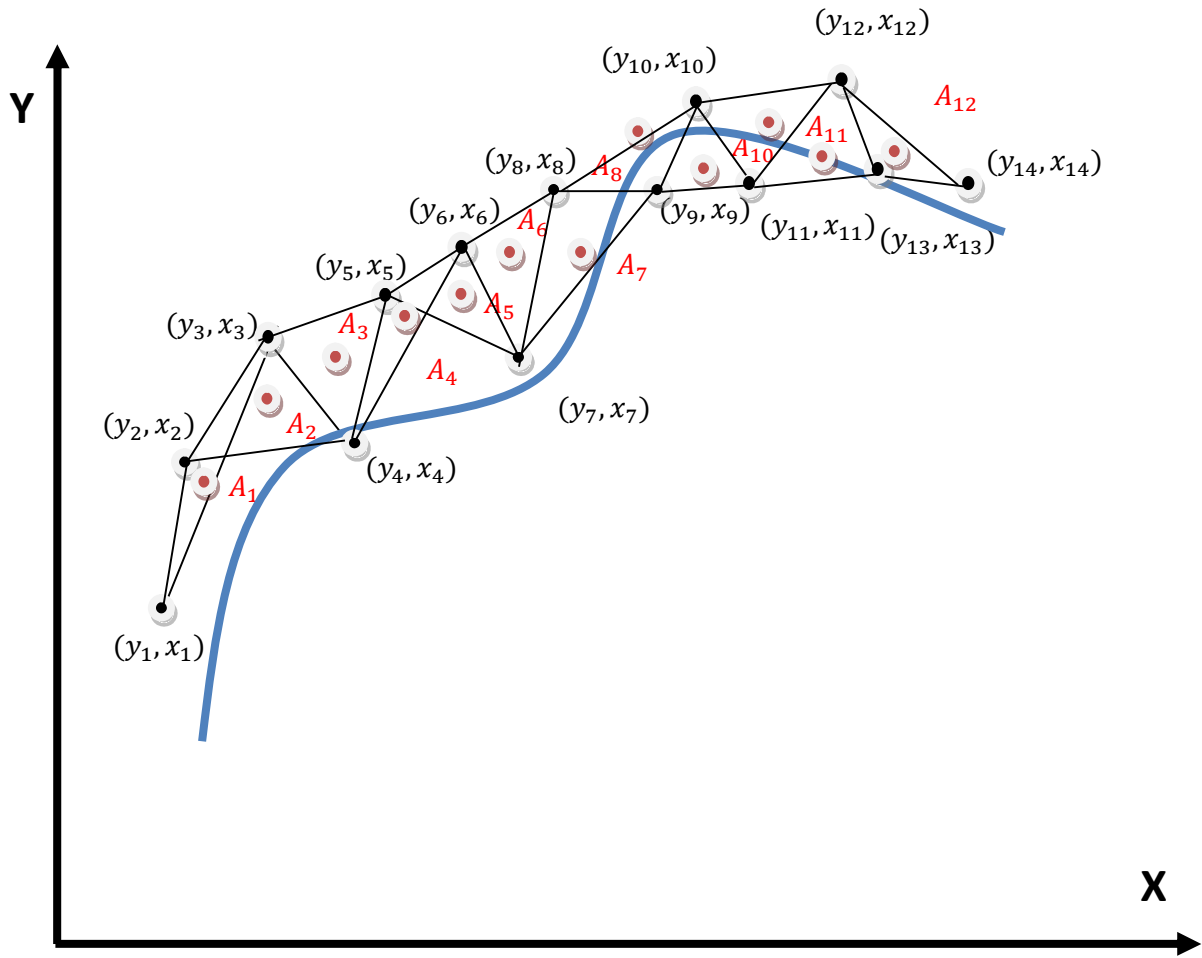

Let's construct triangles whose vertices will lie on the points corresponding to our tabular data

Δ $((x_1, y_1), (x_2, y_2), (x_3, y_3))$

Δ $((x_2, y_2), (x_3, y_3), (x_4, y_4))$

Δ $((x_3, y_3), (x_4, y_4), (x_5, y_5))$

……………………………………….

Δ $((x_{11}, y_{11}), (x_{12}, y_{12}), (x_{13}, y_{13}))$

Δ $((x_{12}, y_{12}), (x_{13}, y_{13}), (x_{14}, y_{14}))$

Find the centroids (the points where the medians intersect) of each triangle

Δ $((x_1, y_1), (x_2, y_2), (x_3, y_3)$   -  A1

Δ $((x_2, y_2), (x_3, y_3), (x_4, y_4)$   -  A2

Δ $((x_3, y_3), (x_4, y_4), (x_5, y_5)$   -  A3

……………………………………….

Δ $((x_{11}, y_{11}), (x_{12}, y_{12}), (x_{13}, y_{13})$  -  A11

Δ $((x_{12}, y_{12}), (x_{13}, y_{13}), (x_{14}, y_{14})$  -  A12

The centroids of the triangles are simultaneously the centers of gravity of these triangles

Thus the points

$A_1, A_2, A_3, A_4, A_5, A_6, A_7, A_8, A_9, A_{10}, A_{11}, A_{12}, A_{13}, A_{14}$ will be a smoothed approximation of our numerical
series
$(x_1, y_1), (x_2, y_2), (x_3, y_3), (x_1, y_1), (x_2, y_2), (x_3, y_3), (x_1, y_1), (x_2, y_2), (x_3, y_3), (x_1, y_1), (x_2, y_2), (x_3, y_3),$

$(x_1, y_1), (x_2, y_2), (x_3, y_3)$

Let's take a concrete example, GDP Hungary 1991-2022

Tabl. 1  GDP  Hungary  1991-2022 [8]

| Year number | 1 | 2 | 3 | 4 | 5 | 6 | 7 | 8 |
|---|---|---|---|---|---|---|---|---|
| Year | 1991 | 1992 | 1993 | 1994 | 1995 | 1996 | 1997 | 1998 |
| GDP, Billion USD | 34,75 | 38,73 | 40,12 | 43,17 | 46,43 | 46,66 | 47,3 | 48,71 |

| Year number | 9 | 10 | 11 | 12 | 13 | 14 | 15 | 16 |
|---|---|---|---|---|---|---|---|---|
| Year | 1999 | 2000 | 2001 | 2002 | 2003 | 2004 | 2005 | 2006 |
| GDP, Billion USD | 48 | 49,66 | 55,66 | 68,33 | 85,33 | 100,66 | 110,66 | 122,66 |

| Year number | 17 | 18 | 19 | 20 | 21 | 22 | 23 | 24 |
|---|---|---|---|---|---|---|---|---|
| Year | 2007 | 2008 | 2009 | 2010 | 2011 | 2012 | 2013 | 2014 |
| GDP, Billion USD | 140,19 | 158,33 | 131,07 | 132,18 | 141,94 | 128,81 | 135,68 | 141,03 |

| Year number | 25 | 26 | 27 | 28 | 29 | 30 | 31 | 32 |
|---|---|---|---|---|---|---|---|---|
| Year | 2015 | 2016 | 2017 | 2018 | 2019 | 2020 | 2021 | 2022 |
| GDP, Billion USD | 125,17 | 128,61 | 143,11 | 160,75 | 164,02 | 157,23 | 182,28 | 178,79 |

We show that the function

Y=(0.0195413177087921-0.0238509487595989i)*exp((0.0890760204993891+2.84113084888033i)*t)+(0.0195413177087987+0.0238509487596027i)*exp((0.0890760204993891-2.84113084888033i)*t)+(0.00571184169812315+0.00817725276894694i)*exp((0.169771000605162+2.34979036975419i)*t)+(0.00571184169812215-0.00817725276894668i)*exp((0.169771000605162-2.34979036975419i)*t)+(0.0316107337540147-0.05937760211869i)*exp((0.0790763425127107+1.99859448050242i)*t)+(0.0316107337540011+0.0593776021186917i)*exp((0.0790763425127107-1.99859448050242i)*t)+(0.00956506041001466-0.181512895422157i)*exp((0.074610789188314+1.62641831734931i)*t)+(0.00956506041001267+0.181512895422141i)*exp((0.074610789188314-1.62641831734931i)*t)+(-0.420119359378276-0.208128205453835i)*exp((0.0752958024756055+1.26430179150643i)*t)+(-0.420119359378279+0.208128205453818i)*exp((0.0752958024756055-1.26430179150643i)*t)+(-4.89462606606935-1.04056590652811i)*exp((-0.0220362770633638+0.587264406873871i)*t)+(-4.89462606606906+1.04056590652809i)*exp((-0.0220362770633638-0.587264406873871i)*t)+(42.5506406866118+1.77635683940025e-15i)*exp((0.0541657791433195+0i)*t)+(1.66084975441488+7.54428165864417i)*exp((0.0351795930091244+0.299849579082453i)*t)+(1.6608497544149-7.54428165864425i)*exp((0.0351795930091244-0.299849579082453i)*t)         (1)

Interpolates a numerical series describing Hungary 's GDP in 1992-2022 , indeed

| Year | Year number | [8] Billion USD | Calculated by the formula (1) Billion USD |
|------|-------------|-----------------|-------------------------------------------|
| 1991 | 0 | | |
| 1992 | 1 | 37.33 | 37.33-7.41753534657954e-15i |
| 1993 | 2 | 40.33 | 40.33+8.47073767158756e-15i |
| 1994 | 3 | 43 | 43+4.16730703448451e-15i |
| 1995 | 4 | 45 | 45-1.59148025121603e-14i |
| 1996 | 5 | 46.33 | 46.33+2.00282561799843e-14i |
| 1997 | 6 | 47 | 47.0000000000001+4.77957618332617e-14i |
| 1998 | 7 | 48 | 48+3.15316981489398e-14i |
| 1999 | 8 | 48 | 48+8.09419271872314e-15i |
| 2000 | 9 | 49.66 | 49.66-8.61440119069117e-15i |
| 2001 | 10 | 55.66 | 55.66+2.28742259697998e-14i |
| 2002 | 11 | 68.33 | 68.3300000000001+4.62026544489059e-14i |
| 2003 | 12 | 85.33 | 85.33-4.03604946194011e-14i |
| 2004 | 13 | 100.66 | 100.66+2.28728074372765e-14i |
| 2005 | 14 | 110.66 | 110.66-5.07769619770173e-14i |
| 2006 | 15 | 122.66 | 122.66+1.7522451935896e-14i |
| 2007 | 16 | 137.66 | 137.66+5.51515515104947e-14i |
| 2008 | 17 | 143 | 143+1.19001123343079e-13i |
| 2009 | 18 | 140.33 | 140.33-3.80679718248545e-15i |
| 2010 | 19 | 134.66 | 134.66+1.6303645075698e-14i |
| 2011 | 20 | 133.66 | 133.66-1.10277318692886e-14i |
| 2012 | 21 | 134.66 | 134.66-2.82901967717864e-14i |
| 2013 | 22 | 134.66 | 134.66+2.03764213397355e-13i |
| 2014 | 23 | 133.66 | 133.66-6.9057002440084e-14i |
| 2015 | 24 | 131.33 | 131.33-8.36338442467804e-14i |
| 2016 | 25 | 132 | 132-6.28757633256485e-14i |
| 2017 | 26 | 143.66 | 143.66+4.99349487991197e-14i |
| 2018 | 27 | 155.66 | 155.66-1.16639357199201e-13i |
| 2019 | 28 | 160.33 | 160.33-1.35477030791677e-14i |
| 2020 | 29 | 167.66 | 167.66+9.35389820338192e-15i |
| 2021 | 30 | 172.33 | 172.33-2.88756448789397e-14i |
| 2022 | 31 | | |

We will present the final result on Fig.1

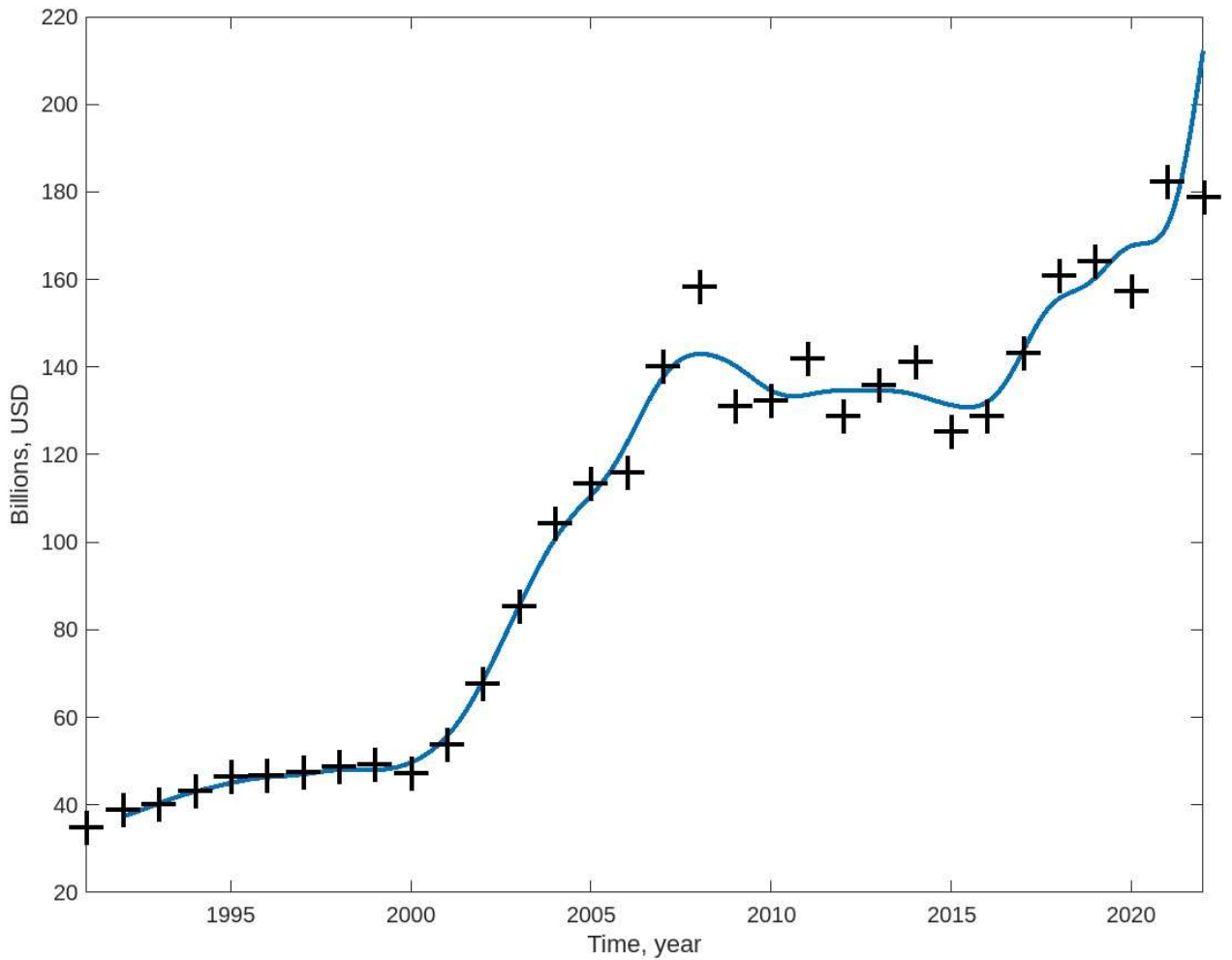

Fig.1 Interpolation of a numerical series describing Hungary's GDP in 1992-2022.
**+** - tabular data [8], ─────── - interpolating function Y (1).

## Discussion

In practice, there is often a need to describe the values given in the form of a table in the form of functional dependence. Usually the observed values have a certain error. For approximation, it is advisable to use a functional dependence, which would make it possible to smooth out the errors of observations at a certain level. In addition, the approximation allows you to determine intermediate values of functions that are not present among the data in the observation table.

Using exponential series for data approximation allows you to get a result no worse than from approximation by polynomials

Approximation in the form of power functions has become widespread in the economic scientific literature. The advantage of this type of approximation is the simple form of the approximating function, and the disadvantage is that in nature not all processes can be described by power functions with a given accuracy. An example is the GDP indicator for several decades . For this case , it is difficult to find a power function approximating a numerical series . But in this case, as shown in this article, you can use exponential series to approximate the data. In this paper, the time series of Hungary's GDP in the period from 1992 to 2022 was approximated by a series of thirty exponents of a complex variable.

## Conclusions

The use of data smoothing by the triangle method allows you to average the data and increase the accuracy of the approximation . This is of practical importance if the observed random variable contains outliers that need to be smoothed out.